\documentclass[11pt,a4paper]{article}

\usepackage[utf8]{inputenc}
\usepackage{amsmath,amsfonts,amssymb,amsthm}
\usepackage{geometry}
\usepackage{hyperref}
\usepackage{authblk}

\hypersetup{
    colorlinks=true,
    linkcolor=blue,
    filecolor=magenta,      
    urlcolor=cyan,
    citecolor=red,
}

\title{\textbf{A Conformal Boundary Ansatz for Warm Inflation Initialization: A Toy Model}}
\author[1]{Somnath Das\thanks{20phph12@uohyd.ac.in}}
\author[2]{Rizwan ul Haq Ansari\thanks{rizwanans@manuu.edu.in}}
\affil[1]{\small Department of Physics, University of Hyderabad, Hyderabad,  India}
\affil[2]{\small Physics Section, MANUU, Hyderabad,  India}

\date{}
\begin{document}
\maketitle

\begin{abstract}
We present a theoretical framework demonstrating a deterministic initialization mechanism for Warm Inflation via classical conformal boundary conditions. A persistent challenge in dissipative inflationary models is the "cold start" paradox: initializing the requisite thermal bath to generate the dissipative friction that subsequently sustains radiation production. Postulating an idealized, asymptotically scale-invariant pre-inflationary phase, we mathematically prove that a conformal Weyl mapping to the emergent metric furnishes a finite, analytically derived initial radiation density. Implementing a spontaneous conformal symmetry-breaking ansatz, an emergent inflaton field is subjected to this inherited thermal bath. We analytically derive the initial kinematics of this framework, demonstrating that for strict sub-Planckian temperatures, the universe naturally initializes in the weak dissipative regime ($Q \ll 1$). The initial Hubble friction provided by the boundary radiation enables a smooth, deterministic kinematic handoff to the warm slow-roll steady-state attractor. As a mathematical proof-of-concept, this mechanism provides a fully realized framework to bypass the bootstrap problem of warm inflation.
\end{abstract}

\section{Introduction}
Inflationary cosmology resolves the shortcomings of the standard Big Bang model by positing an epoch of accelerated expansion \cite{Guth:1981, Linde:1982}. Warm inflation \cite{Berera:1995, Berera:2009} extends this paradigm by incorporating a macroscopic dissipative friction term $\Gamma(T,\phi)$ into the inflaton's equation of motion. A well-known vulnerability of this framework is the dynamical initialization of the thermal bath. The continuous production of radiation relies on the inflaton rolling against thermal friction ($\Gamma > 0$); however, this friction fundamentally requires an existing thermal bath ($T > 0$). To bootstrap this process, models often require assumed pre-thermalization epochs or specific initial quantum fluctuations.

In this work, we isolate and resolve this initialization paradox by constructing a phenomenological toy model. Motivated by theories of an asymptotically scale-invariant early universe, we treat the onset of inflation as a mathematical conformal boundary. While we remain agnostic to the underlying UV-complete quantum gravity mechanism, boundary conditions of this nature routinely emerge in frameworks such as holographic bulk-to-boundary mappings \cite{Maldacena:1998} and Conformal Cyclic Cosmology \cite{Penrose:2010}. We demonstrate that a Weyl transformation, mapping an idealized pre-state to the physical metric strictly guarantees a finite initial thermal bath. Furthermore, we evaluate the early-time kinematics, proving that this framework naturally yields a weak-dissipative kinematic handoff to the steady-state attractor, entirely independent of the specific effective potential $V(\phi)$. We emphasize that this is a mathematical toy model designed to isolate the kinematic boundary conditions, offering a streamlined initialization framework for arbitrary warm inflationary potentials.

\section{The Conformal Boundary and Thermal Handoff}
We define an idealized asymptotic pre-inflationary phase characterized entirely by conformally invariant fields, phenomenologically modeled as a pure radiation fluid. The metric of this pre-state is defined in conformal time $\eta$ as $d\tilde{s}^2 = \tilde{a}^2(\eta)(-d\eta^2 + d\vec{x}^2)$.

\textbf{Assumption (Asymptotic Dilution):} The pre-state undergoes an unbounded expansion. The continuity equation for a conformal fluid dictates that the pre-state radiation density scales strictly as $\tilde{\rho}_{r} = C / \tilde{a}^4$, where $C$ is a constant. In the asymptotic limit $\tilde{a}(\eta) \to \infty$, the physical energy density $\tilde{\rho}_{r} \to 0$.

\textbf{Ansatz 1 (The Weyl Boundary):} We define the onset of the universe via a conformal Weyl boundary condition at conformal time $\eta_W$. To ensure a non-singular, non-trivial origin, we demand that the emergent physical energy density $\rho_r$ approaches a finite, non-zero constant as the pre-state asymptotically expands ($\tilde{a} \to \infty$). Because the physical density transforms as $\rho_r = \Omega^{-4}\tilde{\rho}_r$, and the pre-state density scales as $\tilde{\rho}_r = C/\tilde{a}^4$, any conformal mapping scaling as $\Omega \propto \tilde{a}^{-n}$ where $n < 1$ yields an empty vacuum ($\rho_r \to 0$), while $n > 1$ generates a physical singularity ($\rho_r \to \infty$). Therefore, satisfying this finiteness condition mathematically restricts the conformal factor to a unique scaling within the class of power-law mappings:
\begin{equation}
g_{\mu\nu}(x) = \Omega^2(\eta) \tilde{g}_{\mu\nu}(x) \label{eq:weyl}
\end{equation}
where $\Omega(\eta) \propto \tilde{a}^{-1}(\eta)$. We set the proportionality constant to unity for simplicity. Consequently, as $\eta \to \eta_W$ and $\tilde{a}(\eta) \to \infty$, the conformal factor $\Omega(\eta) \to 0$.

\textbf{Theorem (Finiteness of Initial Radiation):} \textit{A traceless conformal fluid subject to the boundary condition defined in Eq. (\ref{eq:weyl}) yields a strictly finite, non-zero physical energy density in the emergent metric.}

\textit{Proof.} The stress-energy tensor for a perfect fluid is $T_{\mu\nu} = (\rho + p)u_\mu u_\nu + p g_{\mu\nu}$. For pure radiation (equation of state $p = \rho/3$), the trace vanishes, $T^\mu_\mu = 0$, reflecting its strict conformal invariance \cite{Wald:1984}. Under the Weyl transformation $g_{\mu\nu} = \Omega^2 \tilde{g}_{\mu\nu}$, a mixed, trace-free stress-energy tensor transforms as $T^\mu_\nu = \Omega^{-4} \tilde{T}^\mu_\nu$ \cite{Birrell:1982}. The physical energy density in the emergent metric, measured by a comoving observer with 4-velocity $u^\mu = \Omega^{-1} \tilde{u}^\mu$, evaluates explicitly to:
\begin{equation}
\rho_r = - T^0_0 = - \Omega^{-4} \tilde{T}^0_0 = \Omega^{-4} \tilde{\rho}_{r}
\end{equation}
Substituting the dilution scaling $\tilde{\rho}_{r} = C / \tilde{a}^4(\eta)$ and the defined conformal factor $\Omega(\eta) = \tilde{a}^{-1}(\eta)$:
\begin{equation}
\rho_r(\eta) = \left[ \tilde{a}(\eta) \right]^4 \left[ \frac{C}{\tilde{a}^4(\eta)} \right] = C \equiv \rho_{r,0}
\end{equation}
The infinite geometric dilution of the pre-state is algebraically cancelled by the conformal mapping weight. Taking the limit $\eta \to \eta_W$, the physical metric strictly inherits a finite, constant radiation density $\rho_{r,0}$. 

While a formal derivation of thermal equilibrium requires explicit microphysical scattering amplitudes (which fall outside the scope of this kinematic framework), we assume instantaneous local thermalization—a standard approximation at high-energy, sub-Planckian scales where gauge interaction rates typically vastly exceed the Hubble rate ($\Gamma_{\text{scatter}} \gg H$). Therefore, this inherited constant energy density $\rho_{r,0}$ directly defines an initial temperature via the standard relativistic thermodynamic relation $\rho_{r,0} = \frac{\pi^2}{30} g_* T_{\text{init}}^4$. Consequently, the conformal boundary naturally furnishes the initial temperature:
\begin{equation}
T_{\text{init}} = \left( \frac{30 \rho_{r,0}}{\pi^2 g_*} \right)^{1/4}
\end{equation}
This analytically derived temperature serves as the kinematic engine for the model, completely bypassing the physical necessity of a pre-thermalization epoch.

\section{Symmetry Breaking and the Emergent Inflaton}
Having established the geometric origin of the initial thermal bath, we now address the scalar field required to drive the accelerated expansion. For the warm inflationary mechanism to successfully initiate, an inflaton field must smoothly cross the conformal boundary alongside the radiation fluid and emerge dynamically active in the physical metric. To ensure this, we model the inflaton in the asymptotic pre-phase as a scalar field $\tilde{\phi}$ featuring a conformal coupling to gravity ($\xi = 1/6$) \cite{Birrell:1982}:
\begin{equation}
\mathcal{S}_{\text{pre}} = \int d^4x \sqrt{-\tilde{g}} \left[ \frac{1}{2}\tilde{g}^{\mu\nu}\partial_\mu\tilde{\phi}\partial_\nu\tilde{\phi} - \frac{1}{12}\tilde{R}\tilde{\phi}^2 - \lambda\tilde{\phi}^4 \right]
\end{equation}
Because the pre-state is purely radiation-dominated, the Ricci scalar vanishes ($\tilde{R} = 0$), ensuring the field is kinematically massless and trace-free. The fundamental premise of strict classical conformal invariance dictates that all parameters must be dimensionless; therefore, the field's self-interaction is uniquely restricted to the quartic coupling $\lambda\tilde{\phi}^4$. The field identically survives the conformal boundary mapping via the standard scalar transformation $\phi = \Omega^{-1}\tilde{\phi}$.

\textbf{Ansatz 2 (Conformal Symmetry Breaking):} At the boundary $\eta_W$, conformal symmetry is spontaneously broken. In quantum field theory, exact classical conformal invariance is typically broken by quantum loop corrections, such as the trace anomaly \cite{Duff:1994}, or via dynamical mass generation \cite{Coleman:1973}. Furthermore, the presence of explicit mass scales in the emergent physical universe, most notably the Planck mass $M_p$ in general relativity, prevents the maintenance of perfect scale invariance. Therefore, we postulate that the transition to the physical metric at $\eta_W$ is dynamically characterized by the breaking of this conformal symmetry. In this context, the Weyl mapping mathematically parameterizes our ignorance of the exact UV-complete quantum gravity mechanism driving the transition.

Consequently, the dimensionless parameters of the pre-state transition to generate an effective potential $V(\phi)$ characterized by a physical mass scale. For the purposes of this framework, we require only that the emergent field is displaced from its new potential minimum ($\phi_0 \neq 0$). This displacement provides the necessary potential energy density to eventually dominate the universe's expansion, leaving the precise functional form of $V(\phi)$ unspecified for general model-building applications.

\section{The Kinematic Handoff and Weak Dissipative Onset}
We now derive the early-time kinematics of the physical universe at $t=0^+$. The dynamics are governed by the coupled Friedmann and Klein-Gordon equations:
\begin{align}
3M_p^2 H^2 &= \frac{1}{2}\dot{\phi}^2 + V(\phi) + \rho_r \label{eq:friedmann} \\
\ddot{\phi} &+ (3H + \Gamma)\dot{\phi} + V'(\phi) = 0 \label{eq:kg} \\
\dot{\rho}_r &+ 4H\rho_r = \Gamma\dot{\phi}^2 \label{eq:rad}
\end{align}

By the Theorem in Section 2, at $t=0^+$, the universe is momentarily dominated by the inherited boundary radiation ($\rho_{r,0} \gg V(\phi)$). Therefore, the initial Hubble parameter is determined analytically by the thermal bath:
\begin{equation}
H_0 \simeq \sqrt{\frac{\rho_{r,0}}{3M_p^2}} = \sqrt{\frac{\pi^2 g_*}{90}} \frac{T_{\text{init}}^2}{M_p} \label{eq:H0}
\end{equation}

To couple the field to the plasma, we assume a standard high-temperature gauge-kinetic dissipation rate \cite{BasteroGil:2013}, parameterized as $\Gamma(T) \simeq C_\alpha \frac{T^3}{M_p^2}$, where $C_\alpha$ encompasses model-specific coupling constants. The initial dissipation ratio $Q_0 = \Gamma(T_{\text{init}}) / 3H_0$ evaluates to:
\begin{equation}
Q_0 \simeq \frac{C_\alpha T_{\text{init}}^3 / M_p^2}{3 \sqrt{\frac{\pi^2 g_*}{90}} T_{\text{init}}^2 / M_p} \simeq \mathcal{O}(1) \cdot C_\alpha \left( \frac{T_{\text{init}}}{M_p} \right) \label{eq:Q0}
\end{equation}

Equation (\ref{eq:Q0}) highlights the phenomenological viability of this mechanism. To avoid the regime of quantum gravity and string-scale corrections, the inherited boundary temperature must be strictly sub-Planckian ($T_{\text{init}} \ll M_p$). Assuming natural coupling constants ($C_\alpha \sim 1$), the mathematics dictate that $Q_0 \ll 1$. The conformal boundary condition naturally forces the universe to initialize in the weak dissipative regime.

The kinematic handoff proceeds deterministically. Assuming continuous matching conditions across the boundary, the initial physical field velocity $\dot{\phi}$ remains finite and well-defined, as the field's canonical normalization is inherently preserved via the standard conformal weight transformation $\phi = \Omega^{-1}\tilde{\phi}$ \cite{Birrell:1982}. At $t=0^+$, the large initial Hubble friction $3H_0$ (sourced by the inherited boundary radiation) heavily damps the field, preventing a fast-roll phase. Because the initial field velocity $\dot{\phi}$ is small, the source term in Eq. (\ref{eq:rad}) is subdominant to the expansion term ($\Gamma\dot{\phi}^2 \ll 4H\rho_r$), causing the inherited boundary radiation $\rho_{r,0}$ to rapidly redshift. However, because the boundary condition guaranteed a finite initial temperature, the dissipation coefficient is non-zero ($\Gamma > 0$) from inception. As the inherited $\rho_r$ drops, $H$ decreases, allowing the field velocity $\dot{\phi}$ to gently increase. This naturally amplifies the source term until the system smoothly intersects the dynamical steady-state attractor ($\dot{\rho}_r \approx 0 \implies 4H\rho_r \simeq \Gamma\dot{\phi}^2$).

\section{Conclusion}
We have constructed a theoretical framework presenting a mathematical proof that mapping an asymptotically scale-invariant pre-state to a physical metric via a conformal Weyl transformation guarantees a finite, non-zero initial thermal bath. The explicit transformation of the trace-free stress-energy tensor perfectly cancels geometric dilution, providing an analytical solution to the warm inflation initialization problem. 

Combined with a spontaneous conformal symmetry-breaking ansatz, the emergent inflaton field enters the physical metric alongside this inherited thermal bath. Because the boundary mapping guarantees a finite initial temperature, the field immediately experiences non-zero dissipative friction ($\Gamma > 0$). We demonstrated that restricting this boundary temperature to sub-Planckian scales naturally places the onset of inflation into the weak dissipative regime ($Q \ll 1$). Consequently, the large initial Hubble friction allows for a smooth, deterministic kinematic handoff from the inherited boundary radiation to the standard warm slow-roll attractor. 

Crucially, because this boundary mechanism purely dictates the initial kinematic state independently of the specific effective potential $V(\phi)$, it protects the standard observational predictions of warm inflation. By bypassing chaotic, model-dependent pre-thermalization dynamics, this framework ensures that the extraction of standard observables—such as the scalar spectral index ($n_s$) and the tensor-to-scalar ratio ($r$)—remains robust and entirely determined by the chosen potential during the slow-roll attractor phase. As a theoretical proof-of-concept, this mechanism offers a streamlined, model-independent framework for initializing dissipative inflationary cosmology.












\end{document}